\documentclass[superscriptaddress,aps,amsmath,twocolumn,amssymb]{revtex4}

\usepackage{braket}
\usepackage[normalem]{ulem}
\usepackage{amsmath}
\usepackage{amssymb}
\usepackage{gensymb}
\usepackage{amsthm}
\usepackage{dsfont}
\usepackage{graphics}
\usepackage{mathtools}
\usepackage{bm}
\usepackage{epsfig}
\usepackage{algorithm}
\usepackage[noend]{algpseudocode}
\usepackage{xfrac}

\DeclareMathOperator{\tr}{tr}
\newcommand{\Id}{\mathds{1}}

\newtheorem*{thm*}{Theorem}

\usepackage{tabularx}
\newcolumntype{M}[1]{>{\centering\arraybackslash}m{#1}}
\newcolumntype{C}[1]{>{\centering\let\newline\\\arraybackslash\hspace{0pt}}m{#1}}
\usepackage{multirow}
\usepackage{makecell}

\usepackage{dcolumn}
\usepackage{amsfonts}
\usepackage{amssymb}
\usepackage{amsmath}
\usepackage{amsthm}
\usepackage{latexsym}
\usepackage{epsfig}
\usepackage{color}
\usepackage{lscape}
\usepackage{multirow}
\usepackage{braket}
\usepackage{bbm}
\usepackage{mathrsfs}
\usepackage{fancyhdr}
\usepackage{lastpage}
\usepackage{soul}
\usepackage{subfigure}
\usepackage[normalem]{ulem}
\usepackage[hidelinks=true]{hyperref}

\newcommand{\be}{\begin{equation}}
\newcommand{\ee}{\end{equation}}
\newcommand{\ba}{\begin{array}}
\newcommand{\ea}{\end{array}}
\newcommand{\bea}{\begin{eqnarray}}
\newcommand{\eea}{\end{eqnarray}}

\begin{document}

\title{{Fidelity estimation of quantum states on a silicon photonic chip}}

\affiliation{Quantum Engineering Technology Labs, H. H. Wills Physics Laboratory and Department of Electrical \& Electronic Engineering, University of Bristol, BS8 1FD, UK.}
\affiliation{Institute of Photonics and Quantum Sciences (IPAQS), Heriot-Watt University, Edinburgh, United Kingdom}
\affiliation{National Innovation Institute of Defense Technology, AMS, 100071 Beijing, China.}
\affiliation{Quantum Engineering Centre for Doctoral Training, H. H. Wills Physics Laboratory and Department of Electrical \& Electronic Engineering, University of Bristol, Bristol, United Kingdom}
\affiliation{School of Mathematics, University of Bristol, Bristol, United Kingdom.}
\affiliation{Phasecraft Ltd.}
\affiliation{School of Mathematics, University of Bristol, Bristol, United Kingdom.}
\affiliation{These authors contributed equally to this work.}

\author{Sabine Wollmann}
\affiliation{Quantum Engineering Technology Labs, H. H. Wills Physics Laboratory and Department of Electrical \& Electronic Engineering, University of Bristol, BS8 1FD, UK.}
\affiliation{Institute of Photonics and Quantum Sciences (IPAQS), Heriot-Watt University, Edinburgh, United Kingdom}
\affiliation{These authors contributed equally to this work.}

\author{Xiaogang Qiang}
\email{qiangxiaogang@gmail.com}
\affiliation{National Innovation Institute of Defense Technology, AMS, 100071 Beijing, China.}
\thanks{These authors contributed equally to this work.}
\affiliation{These authors contributed equally to this work.}

\author{Sam Pallister}
\affiliation{Quantum Engineering Centre for Doctoral Training, H. H. Wills Physics Laboratory and Department of Electrical \& Electronic Engineering, University of Bristol, Bristol, United Kingdom}
\affiliation{School of Mathematics, University of Bristol, Bristol, United Kingdom.}

\author{Ashley Montanaro}
\affiliation{School of Mathematics, University of Bristol, Bristol, United Kingdom.}
\affiliation{Phasecraft Ltd.}

\author{Noah Linden}
\affiliation{School of Mathematics, University of Bristol, Bristol, United Kingdom.}

\author{Jonathan C.F. Matthews}
\affiliation{Quantum Engineering Technology Labs, H. H. Wills Physics Laboratory and Department of Electrical \& Electronic Engineering, University of Bristol, BS8 1FD, UK.}

\date{\today}

\begin{abstract}
As a measure of the ``closeness'' of two quantum states, fidelity plays a fundamental role in quantum information theory. Fidelity estimation protocols try to strike a balance between information gleaned from an experiment, and the efficiency of its implementation, in terms of the number of states consumed by the protocol. Here we adapt a previously reported optimal state verification protocol (Phys. Rev. Lett. 120, 170502, 2018) for fidelity estimation of two-qubit states.
We demonstrate the protocol experimentally  using a fully-programmable silicon photonic two-qubit chip. Our protocol outputs significantly smaller error bars of its point estimate in comparison with another widely-used estimation protocol, showing a clear step forward in the ability to estimate the fidelity of quantum states produced by a practical device.    

\end{abstract}

\maketitle

\noindent 
Characterising quantum states is an important step in verifiying that desired states are correctly prepared, and output, from quantum information tasks such as quantum teleportation~\cite{Bennett1993}, quantum key
distribution~\cite{Ekert1991,Bennett2014}, and quantum
computation~\cite{Cirac1999,Hayashi2015,Gheorghiu2019}. A crucial step to achieve this is to verify and characterise the quantum states prepared or output from a physical system either as initial input or desired result of a quantum information processing task. Different approaches exist to certify how close the generated quantum states are to desired ones.
The standard approach is to perform quantum state tomography by fully reconstructing the density matrix. However, tomography is computationally hard and time consuming, because the number of parameters to be reconstructed grows exponentially with the size of the system and, thus, is difficult to scale up to large systems. Moreover, a full density matrix reconstruction is not often needed to infer if a physical system has successfully performed a task.\\
This has triggered research interest in alternative statistical approaches such as quantum state verification.
Such protocols rely in general on suitable sequential pass-or-fail measurements applied on individual states. Ref.~\cite{pallister18} proposed an efficient verification protocol with non-adaptive and local projective measurements and constant overhead with regards to optimal global strategies. Extensions of this protocol have been proposed for verifying two-qubit pure states~\cite{wang2019optimal}, bipartite pure states~\cite{yu2019optimal,li2019efficient}, qubit and qudit GHZ states~\cite{li2020optimal} using adaptive approaches assisted by local measurements and classical communications, quantum states in the adversarial scenario~\cite{zhu2019general,zhu2019efficient}, and have been experimentally demonstrated.
Further, it has been observed that these verification protocols could be used for fidelity estimation~\cite{zhu2019general,li2020optimal,li2019efficient}.\\
Here we describe a practical fidelity estimation strategy for two-qubit quantum states. We show that our local verification protocol (LVP) requires only a minimal set of measurements and that it can produce an estimate of the fidelity with significantly reduced error-bars compared to another fidelity estimation protocol.
For the experimental demonstration, we use a fully-programmable two-qubit silicon photonic chip~\cite{qiang2018large}. We compare our strategy to the previously proposed direct fidelity estimation protocol (DFE)~\cite{Flammia2012}.  Further, we compare the capability of these protocols for estimating quantum states with minimal error. Our experimental results showcase the advantages of our protocol for estimating the fidelity of high-quality quantum states.\\
Our fidelity estimation protocol  makes use of the protocol given in~\cite{pallister18} that was designed for a different, but related, task, namely state verification. There, the scenario was as follows. We have a source that emits states $\sigma_i$, where $i$ denotes the trial number from 1 to $n$.
 Each state is promised to be either the target state $\ket\psi$, or at least $\epsilon$ away from $\ket\psi$, defined by $\bra\psi\sigma_i\ket\psi\leq 1-\epsilon$, 
for all $i$.  The verification task is to decide which of these two situations we have, using the minimum number of runs of the experiments $n$. In the verification scenario in~\cite{pallister18}, on each state $\sigma_i$ we make a binary outcome measurement $\{P_j,\Id-P_j\}$. If we receive the outcome associated to $P_j$ on the state $\sigma_i$, the state has ``passed'' and we move on to measuring state  $\sigma_{i+1}$; if we get the outcome associated to $\Id-P_j$, we conclude the state was not $\ket\psi$.

If there were no restrictions on the type of measurements we can perform, it is best to choose the measurement $\{\ket\psi\bra\psi,1-\ket\psi\bra\psi\}$ on every state $\sigma_i$.  But we are interested in the case that we can only make local measurements.  In~\cite{pallister18}, the optimal local measurement strategy was derived in particular in the case of two-qubit states; optimal meaning needing the smallest number $n$ for a given $\epsilon$ and given confidence parameter $\delta$.  As described in~\cite{pallister18}, it turns out that one should choose the $P_j$ from the same distribution for each run of the experiment $\sigma_i$; $P_j$ should be chosen with probability $\mu_j$.  It is convenient to give the verification protocol in terms of what is called a ``strategy'' in~\cite{pallister18}, namely
$\Omega=\sum_j \mu_jP_j$.\\
If we write the state as $\ket{\psi}=\sin\theta\ket{00}+\cos\theta\ket{11}$ for the range $0 < \theta < \frac{\pi}{2}$, $\theta \neq \frac{\pi}{4}$, then, as given in~\cite{pallister18},
\begin{equation}
\Omega = \frac{2-\sin(2\theta)}{4+\sin(2\theta)}P^+_{ZZ} + \frac{2(1+\sin(2\theta))}{3(4+\sin(2\theta))}\sum_{k=1}^3 (\Id - \ket{\phi_k}\bra{\phi_k}),
\end{equation}
where the states $\ket{\phi_k}$ are orthogonal to $\ket{\psi}$ and are written explicitly in the Supplementary Material.

Furthermore, as shown in~\cite{pallister18}, the optimal strategy $\Omega$ has the form 
\begin{equation} \Omega = (1-q)\ket{\psi}\bra{\psi} + q \Id,\label{FormOfOmega}\end{equation}
where $q$, given in the Supplemental Material, is the smallest value corresponding to a local strategy.

We now turn to the subject matter of this paper, namely fidelity estimation.  We again consider $n$ trials of the experiment, and in each trial $i$ a state $\sigma_i$ is generated. Now we are interested in estimating the average fidelity of the states in the experimental runs to $\ket\psi$. But the form of $\Omega$ in Eq.~\ref{FormOfOmega}, makes it clear that performing the same measurements $\{P_j,\Id-P_j\}$ with the relative frequencies $\mu_j$ leads to an estimate of the fidelity; since the fidelity of a state $\rho$ to $\ket\psi$ can be expressed in terms of $\Omega$:
\begin{equation}\label{eq:fid_from_omega}
F = \frac{1}{1-q}\left[\tr(\Omega\rho) - q\right].
\end{equation}
In summary, while the protocol in~\cite{pallister18} was designed for a different (albeit related) purpose, running that protocol provides an estimate for $F$. On a fraction of the trials $\mu_j$, we repeat the measurement $\{P_j,\Id-P_j\}$. We count the total number of times we get the outcome associated to $P_j$ (as opposed to $\Id-P_j$) and then divide this total by $n$.  This provides an estimate of $\tr(\Omega\rho)$ and hence $F$, using Eq.~\ref{eq:fid_from_omega}.

We can already see a key advantage of LVP compared with other fidelity estimation protocols: in the case where $\rho = \ket{\psi}\bra{\psi}$, each measurement $P_j$ accepts with certainty, and hence the fidelity estimate is 1 with certainty. As we will see below, LVP indeed leads to an estimate which out-performs previous work in practice: the error bars in estimating $F$ are substantially better, for a given number of trials $n$.

For the experimental demonstration, the protocol had to be modified to account for experimental practicalities. 
Firstly, LVP is limited to local, projective measurements in a trusted scenario, where our device faithfully reports a measurement outcome~\cite{pallister18}. In an ideal verification protocol, a random projective measurement is chosen for each trial that an adversarial verifier gains the least amount of information possible. Whilst in an experiment, the measurements are typically chosen in advance, hence the discussed verification protocol deviates in this respect from the optimal theoretical strategy in Ref.~\cite{pallister18}.
Moreover, the path encoding in the photonic chip leads to limits in the rank of projectors that can be applied. It is only capable of carrying out rank 1 projective measurements; i.e. applying projectors of the form $\ket{\eta}\bra{\eta}$ for some product state $\ket{\eta}$. Thus each higher rank projector in the optimal strategy must be ``unpacked'' into rank 1 components. If we let the total integration time be $T$, then the optimal protocol is shown in Table~\ref{tab:chip_protocol}.

\begin{table*}[t!]
\centering
\bgroup
\def\arraystretch{1}
\begin{tabular}{|C{3cm}|c|C{3cm}|}
\hline
Measurement setting & Project onto & Integration time\\
\hline
\hline
1 & $\ket{HH}$ & $\frac{T}{4}\left(\frac{2-\sin(2\theta)}{4+\sin(2\theta)}\right)$\\
\hline
2 & $\ket{VV}$ & $\frac{T}{4}\left(\frac{2-\sin(2\theta)}{4+\sin(2\theta)}\right)$ \\
\hline
3 & $\ket{HV}$ & $\frac{T}{4}\left(\frac{2-\sin(2\theta)}{4+\sin(2\theta)}\right)$ \\
\hline
4 & $\ket{VH}$ & $\frac{T}{4}\left(\frac{2-\sin(2\theta)}{4+\sin(2\theta)}\right)$ \\
\hline
\hline
5 & $(\frac{1}{\sqrt{1+\tan\theta}}\ket{H}+\frac{e^{\frac{2\pi i}{3}}}{\sqrt{1+\cot\theta}}\ket{V}) \otimes (\frac{1}{\sqrt{1+\tan\theta}}\ket{H}+\frac{e^{\frac{\pi i}{3}}}{\sqrt{1+\cot\theta}}\ket{V})$  & $\frac{T}{3}\left(\frac{1+\sin(2\theta)}{4+\sin(2\theta)}\right)$\\
\hline
6 & $(\frac{1}{\sqrt{1+\cot\theta}}\ket{H}-\frac{e^{\frac{2\pi i}{3}}}{\sqrt{1+\tan\theta}}\ket{V}) \otimes (\frac{1}{\sqrt{1+\tan\theta}}\ket{H}+\frac{e^{\frac{\pi i}{3}}}{\sqrt{1+\cot\theta}}\ket{V})$ & $\frac{T}{9}\left(\frac{1+\sin(2\theta)}{4+\sin(2\theta)}\right)$ \\
\hline
7 & $(\frac{1}{\sqrt{1+\tan\theta}}\ket{H}+\frac{e^{\frac{2\pi i}{3}}}{\sqrt{1+\cot\theta}}\ket{V}) \otimes (\frac{1}{\sqrt{1+\cot\theta}}\ket{H}-\frac{e^{\frac{\pi i}{3}}}{\sqrt{1+\tan\theta}}\ket{V})$ & $\frac{T}{9}\left(\frac{1+\sin(2\theta)}{4+\sin(2\theta)}\right)$ \\
\hline
8 & $(\frac{1}{\sqrt{1+\cot\theta}}\ket{H}-\frac{e^{\frac{2\pi i}{3}}}{\sqrt{1+\tan\theta}}\ket{V}) \otimes (\frac{1}{\sqrt{1+\cot\theta}}\ket{H}-\frac{e^{\frac{\pi i}{3}}}{\sqrt{1+\tan\theta}}\ket{V})$ & $\frac{T}{9}\left(\frac{1+\sin(2\theta)}{4+\sin(2\theta)}\right)$ \\
\hline
\hline
9 & $(\frac{1}{\sqrt{1+\tan\theta}}\ket{H}+\frac{e^{\frac{4\pi i}{3}}}{\sqrt{1+\cot\theta}}\ket{V}) \otimes (\frac{1}{\sqrt{1+\tan\theta}}\ket{H}+\frac{e^{\frac{5\pi i}{3}}}{\sqrt{1+\cot\theta}}\ket{V})$  & $\frac{T}{3}\left(\frac{1+\sin(2\theta)}{4+\sin(2\theta)}\right)$ \\
\hline
10 & $(\frac{1}{\sqrt{1+\cot\theta}}\ket{H}-\frac{e^{\frac{4\pi i}{3}}}{\sqrt{1+\tan\theta}}\ket{V}) \otimes (\frac{1}{\sqrt{1+\tan\theta}}\ket{H}+\frac{e^{\frac{5\pi i}{3}}}{\sqrt{1+\cot\theta}}\ket{V})$ & $\frac{T}{9}\left(\frac{1+\sin(2\theta)}{4+\sin(2\theta)}\right)$ \\
\hline
11 & $(\frac{1}{\sqrt{1+\tan\theta}}\ket{H}+\frac{e^{\frac{4\pi i}{3}}}{\sqrt{1+\cot\theta}}\ket{V}) \otimes (\frac{1}{\sqrt{1+\cot\theta}}\ket{H}-\frac{e^{\frac{5\pi i}{3}}}{\sqrt{1+\tan\theta}}\ket{V})$ & $\frac{T}{9}\left(\frac{1+\sin(2\theta)}{4+\sin(2\theta)}\right)$ \\
\hline
12 & $(\frac{1}{\sqrt{1+\cot\theta}}\ket{H}-\frac{e^{\frac{4\pi i}{3}}}{\sqrt{1+\tan\theta}}\ket{V}) \otimes (\frac{1}{\sqrt{1+\cot\theta}}\ket{H}-\frac{e^{\frac{5\pi i}{3}}}{\sqrt{1+\tan\theta}}\ket{V})$ & $\frac{T}{9}\left(\frac{1+\sin(2\theta)}{4+\sin(2\theta)}\right)$ \\
\hline
\hline
13 & $(\frac{1}{\sqrt{1+\tan\theta}}\ket{H}+\frac{1}{\sqrt{1+\cot\theta}}\ket{V}) \otimes (\frac{1}{\sqrt{1+\tan\theta}}\ket{H}-\frac{1}{\sqrt{1+\cot\theta}}\ket{V})$  & $\frac{T}{3}\left(\frac{1+\sin(2\theta)}{4+\sin(2\theta)}\right)$\\
\hline
14 & $(\frac{1}{\sqrt{1+\cot\theta}}\ket{H}-\frac{1}{\sqrt{1+\tan\theta}}\ket{V}) \otimes (\frac{1}{\sqrt{1+\tan\theta}}\ket{H}-\frac{1}{\sqrt{1+\cot\theta}}\ket{V})$ & $\frac{T}{9}\left(\frac{1+\sin(2\theta)}{4+\sin(2\theta)}\right)$ \\
\hline
15 & $(\frac{1}{\sqrt{1+\tan\theta}}\ket{H}+\frac{1}{\sqrt{1+\cot\theta}}\ket{V}) \otimes (\frac{1}{\sqrt{1+\cot\theta}}\ket{H}+\frac{1}{\sqrt{1+\tan\theta}}\ket{V})$  & $\frac{T}{9}\left(\frac{1+\sin(2\theta)}{4+\sin(2\theta)}\right)$ \\
\hline
16 & $(\frac{1}{\sqrt{1+\cot\theta}}\ket{H}-\frac{1}{\sqrt{1+\tan\theta}}\ket{V}) \otimes (\frac{1}{\sqrt{1+\cot\theta}}\ket{H}+\frac{1}{\sqrt{1+\tan\theta}}\ket{V})$  & $\frac{T}{9}\left(\frac{1+\sin(2\theta)}{4+\sin(2\theta)}\right)$ \\
\hline
\end{tabular}
\egroup
\caption[The local verification protocol for two-qubit states, decomposed into rank 1 projectors.]{The local verification protocol for the two-qubit state $\protect\ket{\psi} = \sin\theta\protect\ket{00}+\cos\theta\protect\ket{11}$, decomposed into rank 1 projectors. Each block of four settings (1-4, 5-8, 9-12 and 13-16) forms an orthonormal basis; and the integration times sum to $T$.}
\label{tab:chip_protocol}
\end{table*}

Additionally, we consider discrepancies between the theoretical $n\mu_j$ copies of the state $\rho$ to be measured with $P_j$ and the successful recorded data. These are caused by (a) photon loss; (b) variation in counts due to the probabilistic nature of single-photon sources; and (c) fluctuations in the number statistics of quantum states in the experiment.
The errors in the estimated fidelities consist of systematic and statistical errors. The former is result of minor miscalibration and environmental fluctuations, e.g. temperature fluctuations in the lab, which is very hard to determine on such a complex photonic device. The latter stems from Poissonian statistics of the photonic experiment. 
This leads to different run times when performing measurements that form the expectation value. The analysis is discussed in detail in the Supplemental Material (Sec.~\ref{errorbars}).

For the experimental demonstration of the proposed protocol, we use a fully programmable two-qubit silicon photonic chip~\cite{qiang2018large} that generates arbitrary states and performs arbitrary projective measurements. It utilises an optical scheme of linear combination of unitaries based on that an arbitrary two-qubit unitary $U\in \mbox{SU}(4)$ can be decomposed into a linear combination of four terms: $U = \sum_{i=0}^3 {\alpha_i A_i \otimes B_i}$, where $A_i$ and $B_i$ are single-qubit gates and $\alpha_i$ are complex coefficients satisfying $\sum_{i=0}^3 \left|\alpha_i \right|^2=1$, and thus implements universal two-qubit processing via two-photon ququart entanglement~\cite{qiang2018large}. For the photon generation, a continuous-wave 1550~nm laser is amplified and coupled via grating couplers onto the photonic platform. Photon-pairs are then produced in integrated spontaneous four-wave-mixing (SFWM) sources and a path-entangled ququart state $\ket{\Phi}=\alpha_0 \ket{1}_a\ket{1}_e + \alpha_1 \ket{1}_b\ket{1}_f + \alpha_2 \ket{1}_c\ket{1}_g + \alpha_3 \ket{1}_d\ket{1}_h$ is created together with post-selection. Each spatial mode ($a-h$) is further extended into two levels to form qubits $\ket{\varphi_1}$ or $\ket{\varphi_2}$, and $A_i$ and $B_i$ are applied to $\ket{\varphi_1}$ and $\ket{\varphi_2}$ respectively, evolving $\ket{\Phi}$ into $\sum_{i=0}^3 {\alpha_i}A_i \ket{\varphi_1}_{u^i} B_i\ket{\varphi_2}_{v^i}$, where $u^i \in\{a,b,c,d \}$ and $v^i \in \{e,f,g,h \}$. By combining the qubits $a,b,c,d$ into one final-stage qubit, and the qubits $e,f,g,h$ into the second, it yields the state as 
$\sum_{i=0}^3 {\alpha_i A_i \otimes B_i}\ket{\varphi_1}\ket{\varphi_2}$
after path information erasure, realizing the target unitary $U$. Finally the photons are routed off chip via single mode fibres before detection with high-efficiency superconducting single-photon nanowire-photondetectors (SNSPDs) and coincidence counters.
The required two-qubit states, $\ket{\psi} = \sin\theta\ket{00}+\cos\theta\ket{11}$, for this protocol are created by configuring the chip into $\left( \sin{\theta} I \otimes I + \cos{\theta} X \otimes X \right) \ket{0}\ket{0}$, where $I$ and $X$ represent the identity and Pauli-X gates, respectively. 
The projectors in the presented optimal strategy have rank 2 and 3, the chip itself is only capable of measuring rank 1 projectors. Hence each measurement operator must be ``unpacked'' into a lower rank form. See Table~\ref{tab:chip_protocol} for the experimentally implemented measurement settings and corresponding integration times. 

Before the experiment, we perform an initial calibration of our photonic setup. For this, we perform a quantum state tomography of the experimentally generated two-qubit state and use maximum-likelihood estimation to reconstruct the state. Then, we compute the fidelity between the experimental state and the theoretical state, $|\psi\rangle = \sin \theta |00\rangle + \cos \theta |11\rangle$ with $\theta = \frac{k\pi}{32}$ and $k = [0, 16]$. Throughout the experiment, we achieve high-quality quantum states with fidelities of $\geq 90.8\%$ (See Supplemental Material Sec.~\ref{SuppProtocol}).
We perform this calibration step for each value of $k$, before performing the LVP and DFE protocol.

\begin{figure}[htbp]
\centering
\includegraphics[width=0.5\textwidth]{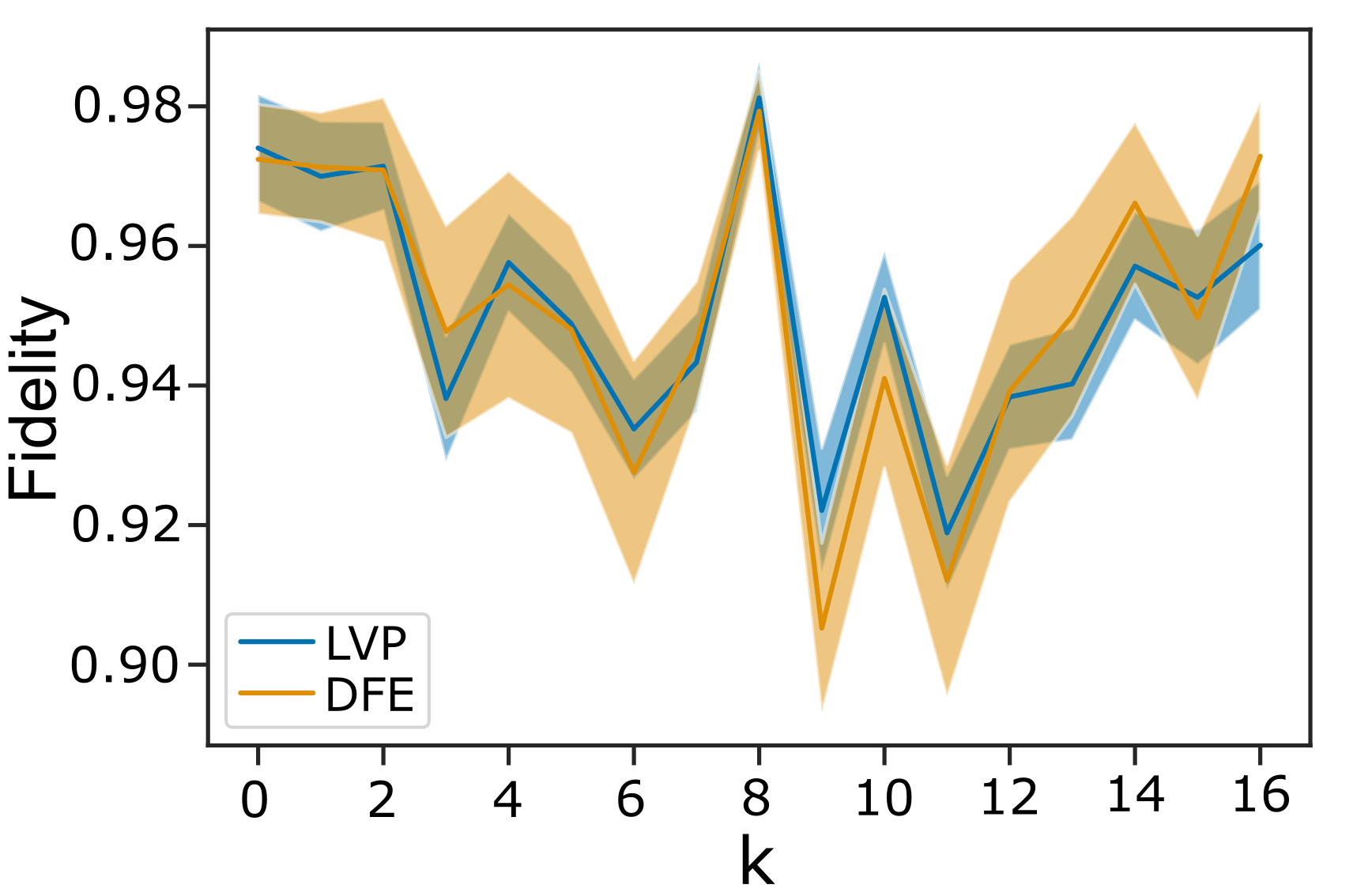}
\caption[A comparison of the optimal fidelity estimation strategy with previous protocols.]{Fidelities and error bars of varying quantum states $|\Psi\rangle = \sin \theta |00\rangle + \cos \theta |11\rangle$ with $\theta = \frac{k\pi}{32}$ and $k = [0, 16]$. We compare our protocol (LVP) against a widely-used direct fidelity estimation (DFE) protocol~\cite{Flammia2012}. 
Both protocols show the same fidelities within error bars across different $|\Psi\rangle$.}\label{FigLOP}
\label{DFEandLVP}
\end{figure}

Then, we experimentally test our fidelity estimation protocols. We generate varying quantum states $|\psi\rangle$ and perform on each of these states the LVP and DFE protocol. For each choice of $k$, the experimental total integration time was fixed at $400$ seconds, giving a total integration time for the whole data set of $\sim 2$ hours. We note that the integration time of the individual measurement setting in this protocol varies with the chosen angle $\theta$ (Table~\ref{tab:chip_protocol}). Calculation of the fidelities and their associated error bars for the entire data set is computationally negligible on a classical computer.
The measured fidelities for LVP are plotted in Fig.~\ref{DFEandLVP} together with the associated error bars. The verified states range between fidelities of $91.9\%$ ($k=12$) to $98.1\%$ ($k=8$) demonstrating consistent high state quality throughout the experiment.

To assess the quality and practicality of the LVP fidelity estimation protocol derived above, we compare the direct fidelity estimation (DFE) protocol~\cite{Flammia2012}. This protocol relies on Pauli measurements to produce an estimate of the fidelity. To give as reasonable a comparison as possible, we integrate for the same total integration time as the protocol derived here~\footnote{The total integration time is divided among 15 Pauli measurement settings, rather than the 4 settings in our protocol.}, and artificially scale the total number of counts in favour of DFE, if they showed significant differences after the same integration time. We make the generous assumption that the DFE was carried out completely perfectly (i.e.\ without error, miscalibration or loss). 
 However, we have to make adaptations to this protocol due to experimental practicalities. Rather than choosing measurement settings at random for each trial, we fix blocks of measurement settings, with size normalised to the expectation of the target state for that particular Pauli observable. The experimental results in Fig.~\ref{FigLOP} show that this protocol produces an estimate with high fidelities for the chosen two-qubit state. The estimates for the LVP and DFE protocols overlap within errors. Furthermore, the error bars on our derived LVP protocol are significantly smaller, showing that our state verification protocol is superior, though the improvement of the error-bars varies with the quantum states.
 
 The rapid technological developments that make the generation of high-quality quantum states rapidly accessible, require protocols for fidelity estimation with minimal uncertainty. Therefore, we investigate the error bars of the LVP and DFE as a function of integration time, thus, the total number of measurements performed.

\begin{figure}[tbp]
\centering
\includegraphics[width=0.48\textwidth]{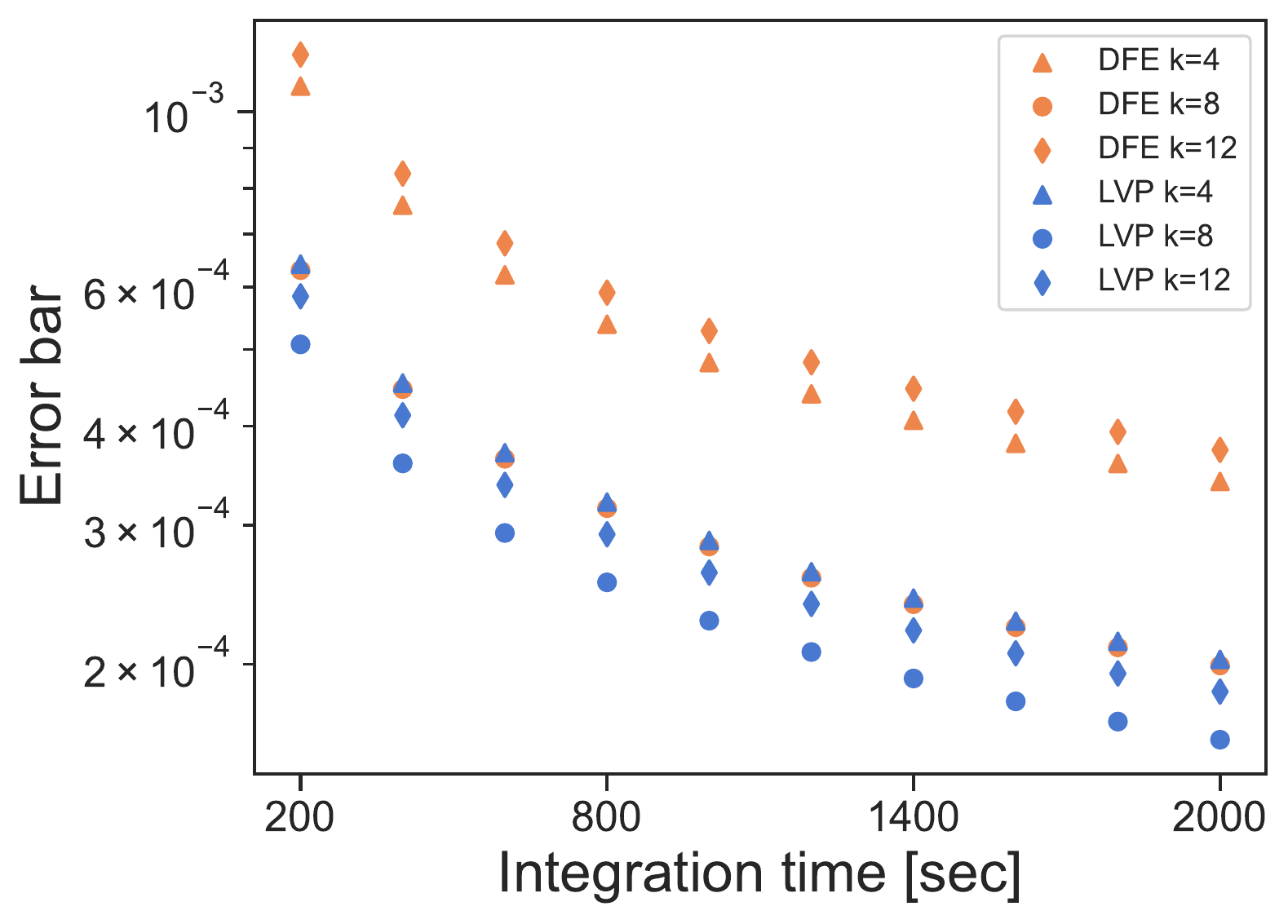}
\caption{A comparison of sizes of error bars for the two fidelity estimation protocols, LVP (blue) and DFE (orange), as a function of integration time. 
We chose $\ket{\psi} = \sin\theta\ket{00}+\cos\theta\ket{11}$ for three choices of $\theta = \frac{k \pi}{32}$ with $k=4, 8, 12$.}
\label{fig:inttimescaling}
\end{figure}
For this, we carried out the LVP protocol for varying integration times, ranging from 200 to 2000 seconds per choice of $\theta$. To compare our results against the DFE, we take the total number of measurements for the original data set integrated over 400 seconds and multiply by a scale factor to get an estimate for the total number of counts to be expected over the same range of integration times. The experimental results are shown in Fig.~\ref{fig:inttimescaling} for our protocol and the DFE taking $\theta \in \left\{\frac{\pi}{8},\frac{\pi}{4},\frac{3\pi}{8}\right\}$. 
For each choice of $\theta$, there is a significant advantage in applying our protocol over the DFE. For $\theta = \frac{12\pi}{32}$ or $\frac{3\pi}{8}$, the error bars for our protocol are a factor of $ \sim 2.03$ smaller than those for DFE. This evidences our proposed protocol is the more optimal strategy when verifying high fidelity quantum states that are realistic scenarios in linear optics experiments.

In conclusion, the protocol we have derived and adapted to experimental practicality represents a clear step forward in the ability to estimate the fidelity of high-quality quantum states generated in an experiment. Broadly, the protocol is advantageous in the following respects: (i) given a fixed number of trials, it outputs significantly smaller error bars on its point estimate than other verification strategies; (ii) it is constructed from non-Pauli measurements and has the ability to alert to coherent measurement errors that are not observed when carrying out ``Pauli'' tomography; and (iii) it requires a negligible amount of post-processing, in contrast with both maximum likelihood and Bayesian tomography~\cite{huszar2012adaptive,granade2016practical}. However, the optimal local fidelity estimation protocol for arbitrary pure states remains an open question. 

\subsection*{Acknowledgements}
We thank Graham Marshall, Xiaoqi Zhou, Jianwei Wang, Callum Wilkes, and Laurent Kling for their contributions to the design and characterisation of the silicon photonic chip.
This project was supported by the Centre for Nanoscience and Quantum Information (NSQI) and the QuantERA ERA-NET Cofund in Quantum Technologies
implemented within the European Union’s Horizon 2020 Programme (QuantAlgo project).
This project has received funding from the European Research Council (ERC) under the European Union's Horizon 2020 research and innovation programme (grant agreement No.\ 817581), and EPSRC grants EP/L015730/1, EP/L021005/1, EP/R043957/1, EP/T001062/1, and EP/M024385/1. SP was supported by the Bristol Quantum Engineering Centre for Doctoral Training, EPSRC grant EP/L015730/1.
Further, we received funding from the European Union’s Horizon 2020
research and innovation programme under the Marie Skłodowska-Curie grant
agreement No 89224. 
X.Q. acknowledges support from National Natural Science Foundation of China (NSFC no. 62075243). 
The data that support the findings of this study are available from the corresponding author upon reasonable request.


\clearpage

\pagebreak
\newpage
\widetext
\begin{center}
\section*{\large Supplementary Material}
\end{center}

\setcounter{equation}{0}
\setcounter{figure}{0}
\setcounter{table}{0}
\setcounter{page}{1}
\setcounter{section}{0}
\setcounter{thm}{0}
\makeatletter
\renewcommand{\theequation}{S\arabic{equation}}
\renewcommand{\thefigure}{S\arabic{figure}}
\renewcommand{\thetable}{S\arabic{table}}
\renewcommand{\bibnumfmt}[1]{[#1]}
\renewcommand{\citenumfont}[1]{#1}

\normalsize

\section{local verification protocol}
The measurement strategy for the local verification protocol (LVP) is $\Omega = \sum_j \mu_j P_j$ for the fidelity estimation with the target state $\ket{\psi}=\sin\theta\ket{00}+\cos\theta\ket{11}$ for the range $0 < \theta < \frac{\pi}{2}$, $\theta \neq \frac{\pi}{4}$, as given in Theorem 1 of Ref.~\cite{pallister18}:
\begin{equation}
\Omega = \frac{2-\sin(2\theta)}{4+\sin(2\theta)}P^+_{ZZ} + \frac{2(1+\sin(2\theta))}{3(4+\sin(2\theta))}\sum_{k=1}^3 (\Id - \ket{\phi_k}\bra{\phi_k}),
\end{equation}
where the states $\ket{\phi_k}$ are
\begin{align}
\ket{\phi_1} &= \left(\frac{1}{\sqrt{1+\tan\theta}}\ket{0} + \frac{e^{\frac{2\pi i}{3}}}{\sqrt{1+\cot\theta}}\ket{1} \right) \otimes \left(\frac{1}{\sqrt{1+\tan\theta}}\ket{0} + \frac{e^{\frac{\pi i}{3}}}{\sqrt{1+\cot\theta}}\ket{1} \right),\\
\ket{\phi_2} &= \left(\frac{1}{\sqrt{1+\tan\theta}}\ket{0} + \frac{e^{\frac{4\pi i}{3}}}{\sqrt{1+\cot\theta}}\ket{1} \right) \otimes \left(\frac{1}{\sqrt{1+\tan\theta}}\ket{0} + \frac{e^{\frac{5\pi i}{3}}}{\sqrt{1+\cot\theta}}\ket{1} \right),\\
\ket{\phi_3} &= \left(\frac{1}{\sqrt{1+\tan\theta}}\ket{0} + \frac{1}{\sqrt{1+\cot\theta}}\ket{1} \right) \otimes \left(\frac{1}{\sqrt{1+\tan\theta}}\ket{0} - \frac{1}{\sqrt{1+\cot\theta}}\ket{1} \right).
\end{align}

\section{Error-bars of the fidelity estimate}\label{errorbars}
 Here, we describe how we calculate the errors for the various protocols.  We use a standard treatment as for example in Ref.~\cite{Taylor}.
 All our expressions are sums of terms of the form
 \[ f= \frac{\sum_\mu a_\mu A_\mu}{\sum_\nu b_\nu A_\nu},
 \]
 where $A_\mu$ is the number of counts received in a channel $\mu$ and $a_\mu$ and $b_\mu$ are constants.

 The squared error associated to this $f$ is
 \begin{eqnarray}(\Delta f)^2&=& \sum_\mu \left(\frac{\partial f}{\partial A_\mu}\right)^2 (\Delta A_\mu)^2\nonumber\\
 &=& \frac{1}{(\sum_\nu b_\nu A_\nu)^2} \sum_\mu \left(a_\mu-b_\mu f\right)^2  A_\mu,\label{general-error-formula}
 \end{eqnarray}
 since
 \[\frac{\partial f}{\partial A_\mu} = \big(a_\mu-b_\mu f\big)\big/\big(\sum_\nu b_\nu A_\nu\big)
 \]
 and
 taking $(\Delta A_\mu)^2=A_\mu$.

\subsection{The locally optimal protocol}
The fidelity is 
\[ F_{LVP} = \frac{\tr(\Omega \rho)-q}{1-q}\]
where
\[
\tr(\Omega \rho)= w_{ZZ} P^+_{ZZ} + \sum_j w_j (1-P(\phi_j)),
\]
where 
\[w_{ZZ}  =\frac{2-\sin(2\theta)}{4+\sin(2\theta)} ,\quad w_j = \frac{2(1+\sin(2\theta))}{3(4+\sin(2\theta))}\quad \forall\, j,\quad q = \frac{2+\sin(2\theta) }{4 +\sin(2\theta)},
\]

and $P(\phi_j)$ is the probability of the measurement of $\{\ket{\phi_j}\bra{\phi_j},\Id - \ket{\phi_j}\bra{\phi_j}\}$ returning the outcome associated to $\ket{\phi_j}$.

We now compute the constants $a_\mu$ and $b_\mu$ for each of the terms in $F_{LVP}$, which will enable us to compute an overall error via (\ref{general-error-formula}). We note that the error in $F_{LVP}$ is the same as that in $\tr(\Omega\rho)/(1-q)$; this is what we compute below.

A feature of the experiments that we are able to perform is that rather than making a measurement and collecting the counts for all the different outcomes for each run, we are only able to collect one outcome at a time.  If we were able to collect all the outcomes, then the expected value of a quantity would simply be computed from the counts for each outcome.  However in our experiments, when we measure, for example,  $Z$ on both qubits, we run the experiment four times, once to collect the outcomes $+$ on each qubit, once $+$ on the first qubit and $-$ on the second and so on.  There may be different run times for each outcome.  We therefore need to take this into account in computing any expectation value.

Consider first the term $w_{ZZ} P^+_{ZZ}$.  Our estimate for $P^+_{ZZ}$ is
\[
\widetilde{P^+_{ZZ}} = \big(\frac{A^{++}_{ZZ}}{r^{++}_{ZZ}}+\frac{A^{--}_{ZZ}}{r^{--}_{ZZ}} \big)\big/ \big(\sum_{\alpha,\beta=+,-} \frac{A^{\alpha\beta}_{ZZ}}{r^{\alpha\beta}_{ZZ}}\big) ,
\]
where $A^{\alpha\beta}_{ZZ}$ is the number of counts in the channel $\alpha\beta$ when we measure $Z$ on both qubits, and $r^{\alpha\beta}_{ZZ}$ is the run time for collecting the $\alpha\beta$ channel.  (Note: we can also take $r^{\alpha\beta}_{ZZ}$ to be proportional to the run time; the formula for the estimator would be the same).

So for measuring $Z$ on both qubits we have 
\[ {a}_{ZZ} = \frac{w_{ZZ}}{1-q}\left( \frac{1}{r^{++}_{ZZ}},0,0,\frac{1}{r^{--}_{ZZ}}\right),\quad
{b}_{ZZ} = \left( \frac{1}{r^{++}_{ZZ}},\frac{1}{r^{+-}_{ZZ}},\frac{1}{r^{-+}_{ZZ}},\frac{1}{r^{--}_{ZZ}}\right).
\]

For the $\phi_j$ measurements we first note that the contribution to the (squared) error in the fidelity $F$ is the same as if we had had $w_jP(\phi_j)$ rather than $w_j(1-P(\phi_j))$.  If we use the notation that $++$ denotes getting the outcome associated to $\ket{\phi_j}$, and $+-,-+,--$ denote the other three outcomes, then for each $j$, the $\bf a$ and $\bf b$ vectors needed in (\ref{general-error-formula}) are
\[ {a}^{(j)} = \frac{w_{j}}{1-q}\left( \frac{1}{r^{++}_{j}},0,0,0\right),\quad
{b}^{(j)} = \left( \frac{1}{r^{++}_{j}},\frac{1}{r^{+-}_{j}},\frac{1}{r^{-+}_{j}},\frac{1}{r^{--}_{j}}\right).
\]
The sum of the squared error is the sum of these squared errors for $ZZ$ and the three different $\phi_j$'s.  The error bars plotted are then the square-roots of this sum of squared errors.

\subsection{Direct fidelity estimation}

The fidelity in this case is $F_{DFE}=\bra\psi \rho \ket \psi$, where $\rho$ is the state produced in the experiment and $\ket\psi$ is the target state $\ket\psi=\sin\theta \ket{00} + \cos\theta \ket{11}$. We can write $\rho$ in its Pauli decomposition:
\[\rho = \frac{1}{4}\sum_{r,s=I,X,Y,Z} \rho_{rs}\Sigma_r\otimes\Sigma_s,
\]
and similarly
\[\ket\psi\bra\psi=\frac{1}{4}\sum_{r,s=I,X,Y,Z} \psi_{rs}\Sigma_r\otimes\Sigma_s,
\]
where $\Sigma_I,\Sigma_X,\Sigma_Y,\Sigma_Z$ are the Pauli operators $I,X,Y,Z$.

Thus \[\rho_{rs} = \tr{\rho\Sigma_r\otimes\Sigma_s}.
\]

Thus, it may be checked that, in terms of $\rho_{rs}$ and $\psi_{rs}$ we may write
\[ F_{DFE} = \frac{1}{4} \big(1+\rho_{XX}\psi_{XX}+\rho_{YY}\psi_{YY} + \rho_{ZZ} +\psi_{IZ}(\rho_{IZ}+\rho_{ZI})\big).
\]
As for the locally optimal protocol, we need to compute the $\bf a$ and $\bf b$ vectors for each of the terms.

Consider first the $\rho_{XX}\psi_{XX}$ term (i.e. corresponding to measuring $X$ on both qubits).

\[ \rho_{XX} = \tr{\rho X\otimes X}= 2 \tr{\rho \Pi^+_{XX}} - 1,
\]
where $\Pi^+_{XX}$ is the projector onto the $+1$ eigenspace of $X\otimes X$.

Now our estimate for $\tr{\rho \Pi^+_{XX}}$ is 
\[
\widetilde{\tr{\rho \Pi^+_{XX}}} = \Big(\frac{A^{++}_{XX}}{r^{++}_{XX}}+\frac{A^{--}_{XX}}{r^{--}_{XX}}\Big)\Big/ \Big(\sum_{\alpha,\beta=+,-} \frac{A^{\alpha\beta}_{XX}}{r^{\alpha\beta}_{XX}}\Big) ,
\]
so, the contribution to the squared error in the fidelity from this term has
\[ {\bf a}_{XX} = \frac{1}{2}\psi_{XX}\left(\frac{1}{r^{++}_{XX}},0,0,\frac{1}{r^{++}_{XX}}\right),\quad
{\bf b}_{XX}  = \left( \frac{1}{r^{++}_{XX}},\frac{1}{r^{+-}_{XX}},\frac{1}{r^{-+}_{XX}},\frac{1}{r^{--}_{XX}}\right).
\]
The $YY$ term is the same except $XX\rightarrow YY$.

Consider finally the term in the fidelity associated to measuring $Z$ on both qubits:
\[f_{ZZ}:= \frac{1}{4} \big( \rho_{ZZ} +\psi_{IZ}(\rho_{IZ}+\rho_{ZI})\big).\]

One can check that the estimates for $\rho_{ZZ},\ \rho_{IZ}$ and $\rho_{ZI}$ are
\begin{eqnarray}
    \widetilde{\rho_{ZZ}} &=& \Big(\frac{A^{++}_{ZZ}}{r^{++}_{ZZ}}-\frac{A^{+-}_{ZZ}}{r^{+-}_{ZZ}}-\frac{A^{-+}_{ZZ}}{r^{-+}_{ZZ}}+\frac{A^{--}_{ZZ}}{r^{--}_{ZZ}}\Big)\Big/  \Big(\sum_{\alpha,\beta=+,-} \frac{A^{\alpha\beta}_{ZZ}}{r^{\alpha\beta}_{ZZ}} \Big),\nonumber\\
\widetilde{\rho_{IZ}} &=& \Big(\frac{A^{++}_{ZZ}}{r^{++}_{ZZ}}-\frac{A^{+-}_{ZZ}}{r^{+-}_{ZZ}}+\frac{A^{-+}_{ZZ}}{r^{-+}_{ZZ}}-\frac{A^{--}_{ZZ}}{r^{--}_{ZZ}}\Big)\Big/  \Big(\sum_{\alpha,\beta=+,-} \frac{A^{\alpha\beta}_{ZZ}}{r^{\alpha\beta}_{ZZ}} \Big),\nonumber\\
\widetilde{\rho_{ZI}} &=& \Big(\frac{A^{++}_{ZZ}}{r^{++}_{ZZ}}+\frac{A^{+-}_{ZZ}}{r^{+-}_{ZZ}}-\frac{A^{-+}_{ZZ}}{r^{-+}_{ZZ}}-\frac{A^{--}_{ZZ}}{r^{--}_{ZZ}}\Big)\Big/  \Big(\sum_{\alpha,\beta=+,-} \frac{A^{\alpha\beta}_{ZZ}}{r^{\alpha\beta}_{ZZ}} \Big),
\end{eqnarray}
where $A^{++}_{ZZ}$ is the number of counts in the $++$ channel and $r^{++}_{ZZ}$ is the run time for collecting $++$, etc.

Thus
\[f_{ZZ}= \frac{1}{4} \Big( \frac{A^{++}_{ZZ}}{r^{++}_{ZZ}}(1+2\psi_{IZ}) -\frac{A^{+-}_{ZZ}}{r^{+-}_{ZZ}}-\frac{A^{-+}_{ZZ}}{r^{-+}_{ZZ}}+
\frac{A^{--}_{ZZ}}{r^{--}_{ZZ}}(1-2\psi_{IZ})\Big)\Big/
\Big(\sum_{\alpha,\beta=+,-} \frac{A^{\alpha\beta}_{ZZ}}{r^{\alpha\beta}_{ZZ}}\Big).
\]
Thus the contribution to the squared error in the fidelity from measuring $Z$ on each qubit has
\[ {\bf a}_{ZZ} = \frac{1}{4}\left(\frac{1+2\psi_{IZ}}{r^{++}_{ZZ}},-\frac{1}{r^{+-}_{ZZ}},-\frac{1}{r^{-+}_{ZZ}},\frac{1-2\psi_{IZ}}{r^{++}_{ZZ}}\right),\quad
{\bf b}_{ZZ}  = \left( \frac{1}{r^{++}_{ZZ}},\frac{1}{r^{+-}_{ZZ}},\frac{1}{r^{-+}_{ZZ}},\frac{1}{r^{--}_{ZZ}}\right).
\]

\section{Quantum State tomography results for calibrating photonic setup}\label{SuppProtocol}

To verify that we experimentally generated the desired state $|\psi\rangle$ in our experiment, we perform a quantum state tomography at every $\theta = \frac{k\pi}{32}$ for an intial calibration of our photonic setup before performing our fidelity estimation protocols. We use maximimum-likelihood estimation to reconstructe the state and then compute the fidelities between our experimental states and the theoretical ones. Table~\ref{fids} shows that we achieve high fidelities of $\geq 90.76\%$ throughout the experiment.

\begin{table}[htbp]
\newcommand{\PreserveBackslash}[1]{\let\temp=\\#1\let\\=\temp}
\newcolumntype{C}[1]{>{\PreserveBackslash\centering}p{#1}}
\newcolumntype{R}[1]{>{\PreserveBackslash\raggedleft}p{#1}}
\renewcommand*{\arraystretch}{1.2}
\begin{tabular}{l|c}
\vspace{5pt}
k  &  Fidelity [\%] \\ 
\toprule
0  & 97.2    \\ \hline
1  & 96.1    \\ \hline
2  & 95.4    \\ \hline
3  & 94.4    \\ \hline
4  & 93.9    \\ \hline
5  & 93.5    \\ \hline
6  & 90.8    \\ \hline
7  & 91.9    \\ \hline
8  & 97.4    \\ \hline
9  & 90.8    \\ \hline
10 & 91.1    \\ \hline
11 & 91.1    \\ \hline
12 & 91.9    \\ \hline
13 & 92.6    \\ \hline
14 & 94.5    \\ \hline
15 & 94.9    \\ \hline
16 & 96.3    \\ \hline
\end{tabular}%
\caption{Fidelities for initial calibration of our photonic setup. We performed a quantum state tomography on our experimentally generated states and computed the fidelity between the reconstructed state and the theoretical state $|\psi\rangle = \sin \theta |00\rangle + \cos \theta |11\rangle$ with $\theta = \frac{k\pi}{32}$ for each $k = [0, 16]$. }\label{fids}
\end{table}

\end{document}